\documentclass[conference]{IEEEtran}
\IEEEoverridecommandlockouts

\usepackage{cite}
\usepackage{amsmath,amssymb,amsfonts}
\usepackage{algorithmic}
\usepackage{graphicx}
\usepackage{textcomp}
\usepackage{xcolor}
\usepackage{comment}
\usepackage{lipsum}
\usepackage{threeparttable}
\def\BibTeX{{\rm B\kern-.05em{\sc i\kern-.025em b}\kern-.08em
    T\kern-.1667em\lower.7ex\hbox{E}\kern-.125emX}}
\begin{document}

\title{Towards Neural Architecture Search for Transfer Learning in 6G Networks}

\author{\IEEEauthorblockN{Adam Orucu\IEEEauthorrefmark{1}\IEEEauthorrefmark{3}, Farnaz Moradi\IEEEauthorrefmark{1}, Masoumeh Ebrahimi\IEEEauthorrefmark{1}\IEEEauthorrefmark{3}, and Andreas Johnsson\IEEEauthorrefmark{1}\IEEEauthorrefmark{2}}
\IEEEauthorblockA{\IEEEauthorrefmark{1} Ericsson Research, Sweden }

\IEEEauthorblockA{\IEEEauthorrefmark{2} Uppsala University, Department of Information Technology, Sweden }

\IEEEauthorblockA{\IEEEauthorrefmark{3} KTH Royal Institute of Technology, Division of Electronics and Embedded Systems,  Stockholm, Sweden}}

\maketitle

\begin{abstract}

The future 6G network is envisioned to be AI-native, and as such, ML models will be pervasive in support of optimizing performance, reducing energy consumption, and in coping with increasing complexity and heterogeneity. A key challenge is automating the process of finding optimal model architectures satisfying stringent requirements stemming from varying tasks, dynamicity and available resources in the infrastructure and deployment positions. In this paper, we describe and review the state-of-the-art in Neural Architecture Search and Transfer Learning and their applicability in networking. Further, we identify open research challenges and set directions with a specific focus on three main requirements with elements unique to the future network, namely combining NAS and TL, multi-objective search, and tabular data. Finally, we outline and discuss both near-term and long-term work ahead.
\end{abstract}

\section{Introduction}

\begin{figure}[t]
  \centering
  \includegraphics[width=.95\linewidth]{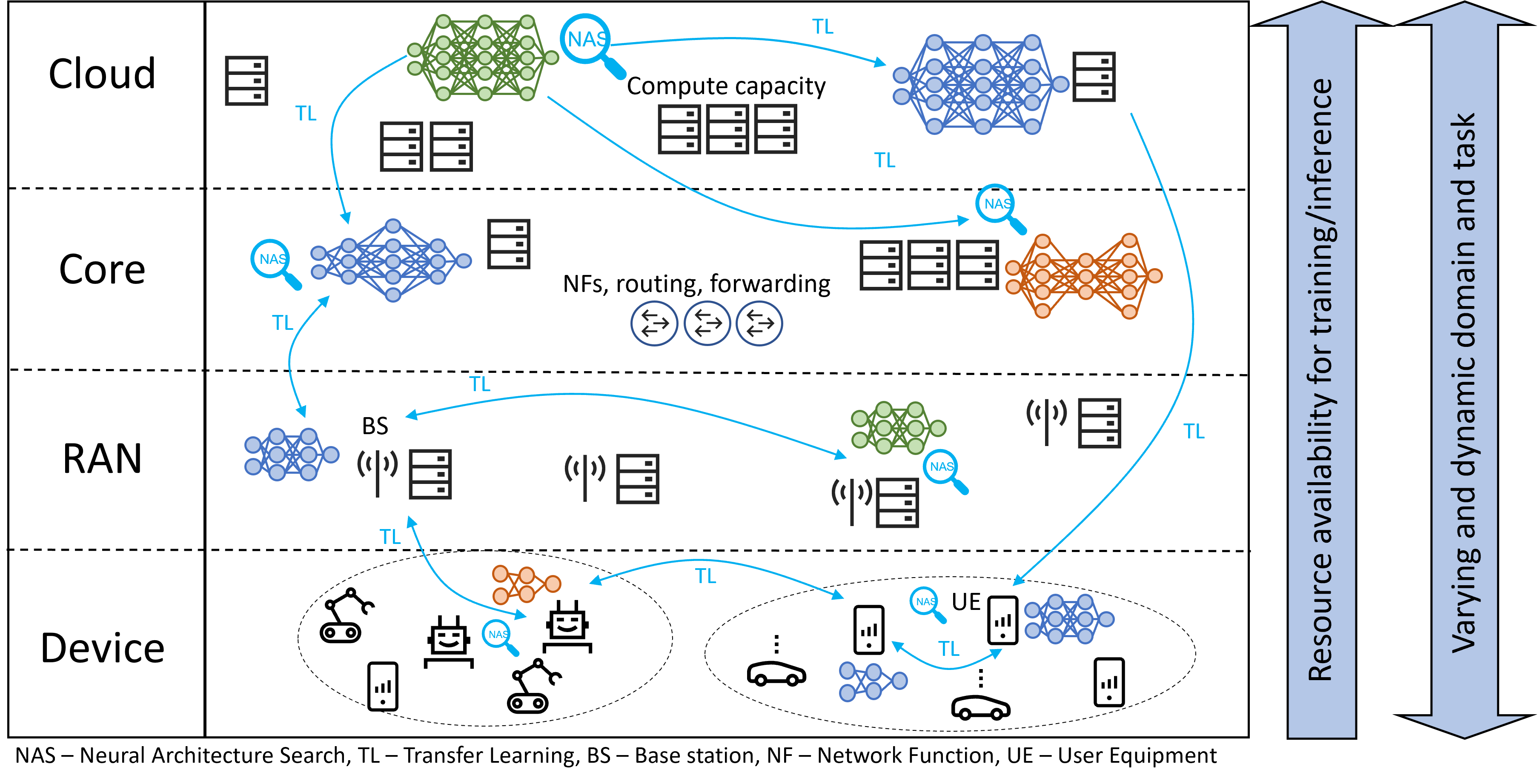}
  \caption{NAS is a key enabler for improved TL and model management in 6G networks having challenging requirements and an abundance of use cases.  }
  \label{fig:concept}
\end{figure}

Artificial Intelligence (AI) and Machine Learning (ML) are technologies which are envisioned to have a prominent role in the future AI-native 6G network. Intelligence and automation will become ubiquitous, and AI/ML will be used to alleviate the increasing complexity and heterogeneity of the network that accommodates a variety of intelligent services and devices~\cite{ainative2023}. 
The number of ML tasks, especially those based on deep learning~\cite{zhang2019deepwireless}, will grow with the heterogeneity of the network, services, and devices. 
This ubiquity is illustrated in Figure \ref{fig:concept}. 
Numerous ML models, with varying resource and data availability and challenging requirements on latency and accuracy, will be deployed in different parts of the network; from devices to base stations,  core network, and central cloud. 

The rapidly increasing number of ML models in the AI-native network necessitates automated ML model life cycle management (LCM). This includes automation of finding optimal model architectures satisfying stringent requirements imposed by different tasks and deployment environments, dynamically adapting to local changes in data and conditions, and automated migration of models.  
The performance of the models can be degraded over time due to dynamic changes in traffic patterns, network reconfiguration, scaling of infrastructure resources, or migration of services between cloud and edge. This dynamicity can result in changes in data distribution, data dimensionality, or even available resources for deployment, causing the previously crafted models to depreciate. This enforces creation of new models from scratch with a new architecture. Training these new models requires collecting labeled data in an operational network which is costly and time-consuming as discussed in~\cite{larsson2023domain}. 

Transfer Learning (TL) is a key technology that can play a significant role in adapting to the changes and heterogeneity in the network and maintaining the high expectations in operational networks~\cite{nguyen2022transfer}. TL techniques can reduce the need for collecting large labeled datasets, as well as the training time, while potentially enhancing models' performance. However, TL methods do not address the challenges associated with automatically finding the best model architecture given the changes in the model requirements such as constrained resources or stricter latency requirements. 

One family of approaches enabling automatic search for the best model architecture for a given task and dataset is Neural Architecture Search (NAS). It is a subfield of automated machine learning (AutoML) and  has received considerable attention in the AI/ML community, particularly for image and language data~\cite{he2021automl}. However, using NAS as an enabler for efficient TL in networking has not been sufficiently addressed in the literature.
Thus, in this paper, we outline research directions to fill this gap. More specifically, our contributions are (1) a comprehensive study of the NAS for TL approaches, and (2) identification of the research challenges of using NAS techniques for improving TL and model LCM for networking.

\section{Background on NAS and TL}
In this section, we provide a background on TL and NAS, 
and review some works on the integration of NAS and TL.

\subsection{Transfer Learning}
Transfer learning aims to use knowledge from existing models to train and fine-tune a new model. That is, transferring the knowledge from an existing model rather than training a new model from scratch.

TL methods are often used when there is insufficient training data or compute budget to train an entire neural network, therefore a network trained on ``similar'' data is used as a starting point for the training. The model the knowledge is transferred from is called source model and is said to be from the \textit{source domain}, while the model intended to be learned and its data are called target model and \textit{target domain}, respectively. The source and target domains differ in their data distribution or the task that the model aims to carry out. However, the source model is still useful for the target domain since there is an overlap in the information contained in both domains.

Transfer learning has received considerable attention in the literature, with
 multiple survey papers published~\cite{pan2009survey, Zhang2022survey}. 
A survey of TL for wireless networks is presented in~\cite{nguyen2022transfer} which identified applications of TL in 5G networks. 
Several recent studies have shown that TL can outperform traditional methods when limited data is available for telecom problems including service performance prediction in dynamically changing environments~\cite{moradi2019performance}, prediction of channel quality across different wireless channels~\cite{Parera2019}, 
antenna tilt configuration~\cite{Parera2020tilt}, prediction of network key performance indicators~\cite{Parera2021}, and downlink beamforming adaptation~\cite{Yuan2021beamforming}. 
TL is also expected to play an important role in 6G networks. A review of the algorithms and future research directions for TL in 6G are discussed in \cite{wang2021transfer}.

\subsection{Neural Architecture Search}

\begin{figure}[t]
  \centering
  \includegraphics[width=.8\linewidth]{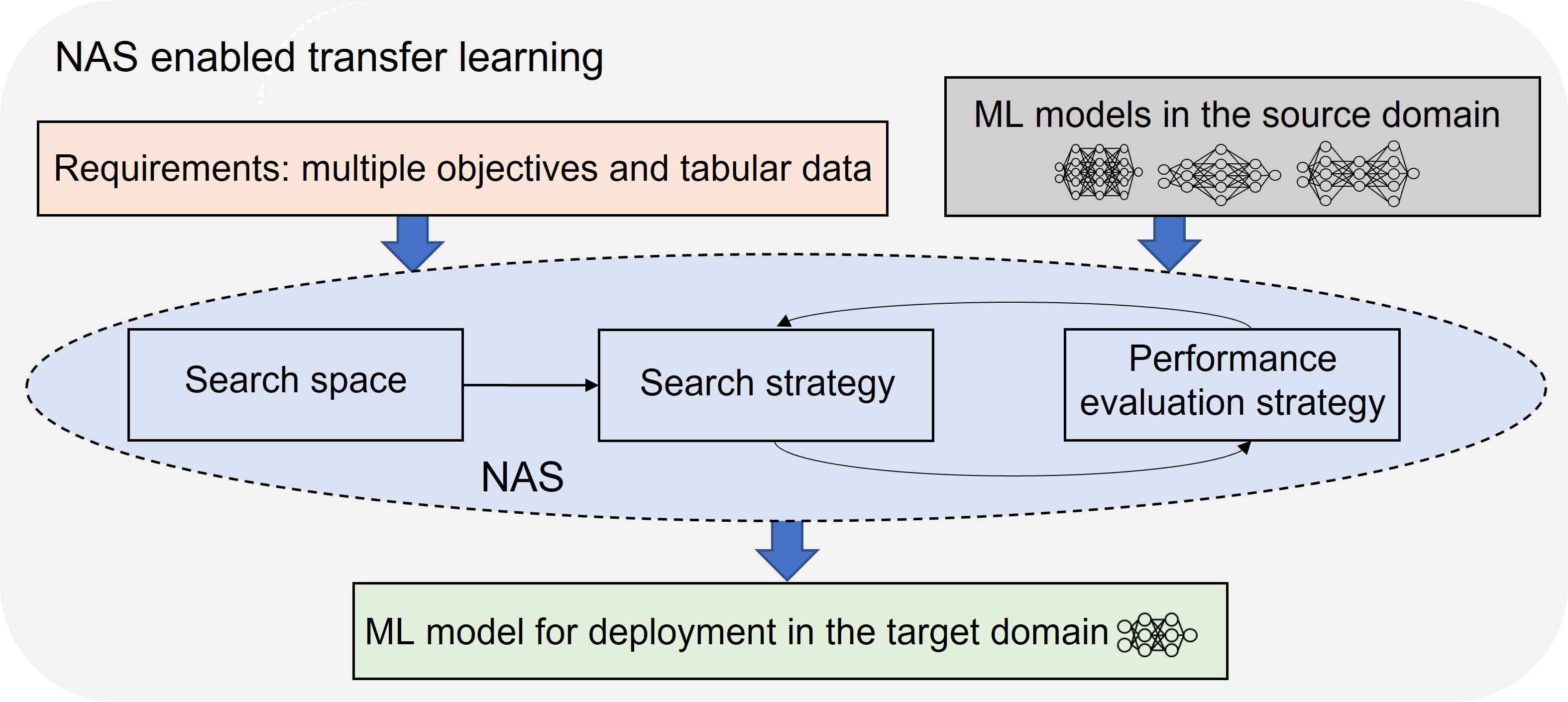}
  \caption{NAS is an enabler for improved transfer of models between domains.}
  \vspace{-0.1in}
  \label{fig:nas}
\end{figure}

The aim of NAS is to find the optimal model architecture for a given task on a specific dataset. As illustrated in Figure~\ref{fig:nas}, the entire process of finding an architecture can be divided into three parts: defining a search space, sampling an architecture through some search strategy, and finally evaluating the sampled architecture. After a search space is defined the sampling and evaluation steps are repeated in a loop until a satisfactory architecture is found or a predefined iteration count is reached. This loop enables the search strategy to obtain feedback from the evaluation to learn the attributes of good architectures. In the following, we discuss these three aspects of NAS.

\subsubsection{Search space}\label{sec:search-space} A baseline for defining the search space is to select allowed architectural parameters such as type of layers and their attributes. 
Then, the search algorithm samples the model architecture from this search space. However, one of the most important challenges for NAS algorithms is the time it takes to find a suitable architecture out of the many possible combinations available through the search space. It is well-known that the search time can be in the order of days. For this reason, many approaches try to further restrict the search space. By far the most common approach, NASNet~\cite{zophLearningTransferableArchitectures2018}, is to let the NAS pick blocks of several layers (known as cells) and then repeat these blocks to form the neural architecture. This significantly shrinks the search space and reduces the search time. 
Other works have additionally restricted the search space by including constraints from the hardware~\cite{benmezianeComprehensiveSurveyHardwareAware2021}.

\subsubsection{Search strategy}
It is important for the methods to be able to learn from previous selections. For this reason, advanced ML solutions are often used during this process. One common approach is to encode the architecture as a string and then to use evolutionary search (ES) to find suitable variations~\cite{elskenEfficientMultiobjectiveNeural2019, realRegularizedEvolutionImage2019, realLargeScaleEvolutionImage2017, liuHierarchicalRepresentationsEfficient2018, luNSGANetNeuralArchitecture2019}. In this process, generally, a starting population is randomly sampled and evaluated. Then in a loop, the best-performing architectures are used to create a new generation which is further evaluated. Another popular approach is to use reinforcement learning (RL) algorithms. In this setting, the selection of an architecture and its performance could be thought of as an agent's action and its reward, respectively. There are numerous work that use RL as the search algorithm in NAS~\cite{zophNeuralArchitectureSearch2017}. 
Finally, other search approaches might use Bayesian Optimization (BO)\cite{luNSGANetNeuralArchitecture2019} or aim to change the search space into a differentiable form and apply commonly used gradient optimization approaches~\cite{liuDARTSDifferentiableArchitecture2019}.

\subsubsection{Performance evaluation} The third part of the NAS process evaluates the performance of an architecture that has been chosen by the search strategy. Since this usually requires training a fresh architecture and then testing, it is the part that is most compute-intensive and is the reason NAS algorithms take a long time to run. Therefore, it is a crucial step to be optimized. The first major improvement in this aspect has been provided by \cite{phamEfficientNeuralArchitecture2018}. The authors created a large computational graph (known as SuperNet) that represents the entire search space and contains every possible architecture as a subgraph (known as a subnet). 
Every time an architecture is chosen by the search algorithm to be evaluated, the corresponding subnet is selected from the SuperNet and extracted including its weights. After these weights are trained and tested, their new values are updated in the SuperNet. This way, if a weight has been trained during a previous selection it already contains some information relevant to the task. By using this method, the number of epochs to train each model is reduced since many of the models already have some knowledge of the task before the training begins. This approach had provided 1000x improvement in architecture search time compared to first methods, and many methods have continued to use SuperNet based approaches. Other approaches for decreasing the evaluation time include learning curve extrapolation methods~\cite{baker2017accelerating}, evaluation on a simpler domain rather than the target dataset~\cite{pandaNASTransferAnalyzingArchitecture2021}, and training surrogate models to predict the performance of the network given its architecture~\cite{luNeuralArchitectureTransfer2021}.

\subsection{NAS and TL}
\label{sec:nas4tl}

Research on approaches for combining NAS and TL methods is also ongoing, and the goal of these approaches is to speed up the NAS process using TL, or creating architectures that work well across multiple domains. In~\cite{zophLearningTransferableArchitectures2018} and \cite{wistubaXferNASTransferNeural2020} the authors propose to use proxy datasets for architecture search. These datasets are similar but less complex than the initial one and are used to find an architecture that would also work well on the original problem, therefore speeding up the search process. A survey of methods using such approaches can be found in \cite{pandaNASTransferAnalyzingArchitecture2021}. ModuleNet~\cite{chen2021modulenet}, on the other hand, takes a different approach. It uses expert-defined trained architectures as the search space, and uses evolutionary search to find a new architecture that uses blocks of layers from different trained architectures including their trained weights. This not only significantly reduces the search space but, more relevantly, transfers the existing knowledge to the new domain.

\section{Challenges and research directions}
In this section, we outline a set of research challenges and directions by identifying gaps in the state-of-the-art. We envision an AI-native 6G network where models can be trained, transferred, and deployed as illustrated in Figure \ref{fig:concept}. More specifically, a source domain ML model is to be transferred to a target domain as depicted in Figure~\ref{fig:nas}. However, the transfer must also consider multi-objective requirements and challenges with tabular data. 
In Table~\ref{tab:example} we highlight state-of-the-art and whether they address these challenges. To this end, we believe that new NAS technology for TL must be researched and developed.

\begin{table}[t]
\centering
\caption{Comparison of relevant NAS approaches.} 
\begin{threeparttable}
\begin{tabular}{|c|c|c|c|c|}
\hline
Method & Search  & Tabular & MO\tnote{1} & TL \\

\hline
\hline

AgEBO~\cite{egeleAgEBOTabularJointNeural2021} & ES \& BO & \checkmark &  &  \\ 
NAT~\cite{luNeuralArchitectureTransfer2021} & ES &  & \checkmark & $\bullet$ \\ 
Once-for-All~\cite{caiOnceforAllTrainOne2020} & N/A\tnote{2} &  & \checkmark & $\bullet$ \\ 
MnasNet~\cite{tanMnasNetPlatformAwareNeural2019} & RL &  & \checkmark &  \\ 
XferNAS~\cite{wistubaXferNASTransferNeural2020} & NAO &  &  & $\bullet$ \\ 
LEMONADE~\cite{elskenEfficientMultiobjectiveNeural2019} & ES &  & \checkmark &  \\ 
TabNAS~\cite{yangTabNASRejectionSampling2022} & RL & \checkmark & \checkmark &  \\ 
Zoph et al.~\cite{zophLearningTransferableArchitectures2018} & RL &  &  & $\bullet$ \\ 
ModuleNet~\cite{chenModuleNetKnowledgeInheritedNeural2022} & ES &  & \checkmark & $\bullet$ \\ \hline\hline
Identified research directions & tbd & \checkmark & \checkmark & \checkmark \\ \hline

\end{tabular}
\begin{tablenotes}
       \item \footnotesize{1: Multi-objective}, \footnotesize{2: No search performed}, $\bullet$: Not fully covered
       
     \end{tablenotes}
     \end{threeparttable}

\label{tab:example}
\end{table}

\subsection{TL challenge}
\label{sec:tl-chal}
In networking and telecom infrastructures the environment evolves throughout its lifetime which could lead to model performance degradation as time passes. In order to prevent degrading and take advantage of previously trained models, TL techniques have been used to adapt ML models to the new environment or task \cite{moradi2019performance}.

However, when transferring a source model to a new target domain using standard TL, the architecture of the model typically doesn't change and stays similar. Of course, this architecture may not necessarily be the one that is optimal for the target domain, since feature space or tasks may be different, there may be limitations in available data, or resource restrictions with respect to compute or memory. Therefore, using novel NAS techniques in the process of transferring the knowledge to the new domain will help creating specialized, effective, and efficient architectures. As described in Section~\ref{sec:nas4tl}, approaches that have jointly used NAS and TL have been studied before, however, the existing methods do not address this particular aspect as indicated in Table \ref{tab:example}. This establishes our first and perhaps the most important challenge; \textit{can we leverage knowledge from a similar source domain in the process of  finding an optimal model architecture for a target domain?}

\subsection{Multi-objective challenge}

The prediction performance of an ML model is indeed not the only aspect to consider when designing a neural network. This is especially true in networking applications where ML models are deployed in a heterogeneous network (as shown in Figure~\ref{fig:concept}). The availability of resources for model training and inference varies across different parts of the network,
such as available memory, computational capacity, and power budget in devices compared to RAN, network core, or centralized cloud. Added limitations lead to additional objectives that must be addressed in the search for a model architecture. Optimization with such goals is called multi-objective optimization.

There is a significant amount of work on this topic relating to NAS, focusing on approaches covering different aspects of the multi-objective challenge. 
Works such as MnasNet~\cite{tanMnasNetPlatformAwareNeural2019}, Once-for-All\cite{caiOnceforAllTrainOne2020}, and LEMONADE~\cite{elskenEfficientMultiobjectiveNeural2019} focus on the accuracy-latency trade-off and try to find the Pareto-optimal solution. 
TabNAS~\cite{yangTabNASRejectionSampling2022} puts a hard cap on the total number of units in the neural network, while NAT~\cite{luNeuralArchitectureTransfer2021} aims to create a SuperNet whose subnets are optimized for various different tasks. Yin et al.~\cite{yinDynamicDataCollection2022} try to find the optimal neural network while restricting the amount of data it can receive during training from edge devices. Hence, our second challenge to address is \textit{how to incorporate multiple objectives stemming from the dynamicity in 6G networks to automatically find the best architecture during transfer?}

\subsection{Tabular data challenge}
An overwhelming majority of the published work on NAS are on image data~\cite{zophNeuralArchitectureSearch2017, liuDARTSDifferentiableArchitecture2019}, while limited efforts have been made considering unstructured and tabular data, which is quite common in networking~\cite{zhang2019deepwireless}. 
Although applying AutoML methods to tabular data has been explored extensively, this is not the case for NAS. According to Yang et al.~\cite{yangTabNASRejectionSampling2022} this is due to lack of understanding of promising architectures and lack of relevant datasets. This creates a research opportunity in applications of NAS and TL for tabular data. Although few, there still is some work in the area. The most noteworthy ones are AgEBO-Tabular~\cite{egeleAgEBOTabularJointNeural2021} and TabNAS~\cite{yangTabNASRejectionSampling2022}. They, respectively, use genetic algorithms and reinforcement learning to find the best fully-connected neural networks for tabular data. Our final challenge is then \textit{how to improve and adapt NAS methods for TL with specific focus on tabular networking datasets?}

\section{Ongoing and Planned Work}
The main goal of our research is to develop and use the NAS technology as an enabler for automatic LCM in AI-native 6G networks by improving  
transfer of knowledge for different ML models. Our approach, as depicted in Figure~\ref{fig:nas}, will be to incorporate the requirements from the network as well as the knowledge from existing ML models for relevant tasks into the architecture search in order to identify the optimal model architecture to be deployed in the target domain. We are currently (1) investigating selected use cases and associated requirements for 6G networks. We will (2) develop a new search space that is more suitable for multi-objective tabular data based on knowledge from the source domain, (3) propose an evaluation metric that will enable us to represent multiple objectives, such as latency or power requirements alongside model performance, and (4) define a search strategy that will combine these procedures with models available to transfer from the source domain.

\section*{Acknowledgements}
{This research was partially supported by the Wallenberg AI, Autonomous Systems and Software Program (WASP) funded by the Knut and Alice Wallenberg Foundation.

\bibliographystyle{IEEEtran}
\bibliography{references}
\end{document}